\documentclass{ws-ijmpe}

\begin{document}

\markboth{Authors' Names}{Instructions for
Typing Manuscripts (Paper's Title)}

\catchline{}{}{}{}{}
%%%%%%%%%%%%%%%%%%%%%%%%%%%%%%%%%%%%%%%%%%%%%%%%%%%%%%%%%%%%%%%%%%%%

\title{MASSES AND RADII OF THE NUCLEI WITH $N \geq Z$ IN AN ALPHA-CLUSTER MODEL\\}

\author{\footnotesize G.K. NIE\footnote{galani@Uzsci.net}}

\address{Institute of Nuclear Physics, Uzbekistan Academy of Sciences,
100214 Tashkent, Uzbekistan}

\maketitle

\begin{history}
\received{(received date)}
\revised{(revised date)}
%\accepted{(Day Month Year)}
%\comby{(xxxxxxxxxx)}
\end{history}

\begin{abstract}
In the framework of a recently developed alpha-cluster model a nucleus is represented as a core (alpha-cluster liquid drop with dissolved excess neutron pairs in it) and a nuclear molecule on its surface. From analysis of experimental nuclear binding energies one can find the number of alpha-clusters in the molecule and calculate the nuclear charge radii. It was shown that for isotopes of one $Z$ with growing $A$ the number of alpha-clusters in the molecule decreases to three, which corresponds to the nucleus $^{12}$C for even Z and $^{15}$N for odd $Z$, and the specific density of the core binding energy $\rho$ grows and reaches its saturation value.
In this paper it is shown that the value $\rho$=2.55 MeV/fm$^3$ explains the particular numbers of excess neutrons in stable nuclei.
\end{abstract}

\section{Introduction}

In the framework of the alpha-cluster model \cite{1} an atomic nucleus is represented as a core and a nuclear molecule on its surface (the notion of a nuclear molecule on the surface of a nucleus was developed in \cite{2}). Let us consider the nuclei $A(Z,N)$ and $A_1(Z+1,N+1+1)$, $Z$ and $N$ are even numbers. The nucleus $A_1$ in comparison with $A$  has an additional single $pn$-pair with one excess neutron glued to it. (These nuclei are selected due to the well known fact that the stable nuclei with even $Z$ in most of the cases have even number of excess neutrons $\Delta N=N-Z$ and with odd $Z_1=Z+1$ the value $\Delta N=N+2-Z_1$ is odd). Then in the framework of the model the outside molecules in $A$ and $A_1$ nuclei consist of $N_\alpha^{ml}$ and  $N_\alpha^{ml}$+0.5 $\alpha$-clusters correspondingly. The cores of these nuclei consist of the equal numbers of $\alpha$-clusters $N_{\alpha}^{core}=N_\alpha -N^{ml}_\alpha$, where $N_\alpha=Z/2$, and of equal numbers of excess $nn$-pairs $N_{nn}=(N-Z)/2$. The validity of this representation is approved by the fact that the separation energies of $nn$-pairs in these two nuclei $A$ and $A_1$ in most of the cases are equal (with an accuracy in 1 MeV).

The Coulomb energy, the energy of nuclear force of the links between neighboring clusters and the surface tension energy of the core are calculated on the number $N_{\alpha}^{core}$. The binding energy of the excess $nn$-pairs is calculated in dependence on the number of the pairs $N_{nn}$. It makes the number of $\alpha$-clusters outside the core $N^{ml}_\alpha$ the only unknown parameter in calculation of nuclear binding energy, which allows one to find the number from fitting experimental binding energy. The analysis shows\cite{1} that in isotopes of one $Z$ with growing $A$ the number $N^{ml}_\alpha$ decreases in most of the cases to $N^{ml}_\alpha$=3 and the specific density of core binding energy $\rho $ slowly increases to reach its saturation value in stable and beta-stable isotopes $\rho=2.60\pm$ 0.1 MeV/fm$^3$.
Obtained values $N^{ml}_\alpha$ also allow one to calculate the radii\cite{1}. Besides, it gives an explanation of the fact that the experimental charge radii of the isotopes belonging to one $Z$ decrease with growing $A$.

In this work some new proofs of the model have been found. It is shown here that the binding energies of the chain of the stable nuclei $A(Z,N)$ and $A_1(Z_1,N+2)$ (there is only one rare two stable isotopes with odd $Z_1$, therefore the chain is defined by the actual numbers of the $A_1$) are described by a simple formula with the only fitting parameter $\rho $=2.55 MeV/fm$^3$ in suggestion that a stable nucleus consists of a core and the molecule $^{12}$C ($^{15}$N in case  $A_1(Z_1,N+2)$) outside the core.  Besides, the constant value $\rho $ provides an explanation of the particular number of excess neutrons in stable nuclei.

\section{Main Formulas}
\subsection{Representation of nucleus as a liquid $\alpha$-cluster drop}
The first step in developing the phenomenological alpha-cluster model was the
finding that the binding energies $E_{N_\alpha}$ of the symmetrical nuclei with $Z\leq 28$ $(\Delta N=0)$ are described by the simple formula\cite{3} (here energies are given in MeV and radii in fm)
\begin{equation}
E_{N_\alpha}=N_{\alpha }\varepsilon_{\alpha }+3(N_{\alpha }-2)\varepsilon _{\alpha \alpha },
\label{1}
\end{equation}
where $\varepsilon _{\alpha }=E_{^{4}\mathrm{He}}=28.296$ MeV, $\varepsilon _{\alpha \alpha }=2.425$ MeV, and for the odd $Z_1$ nuclei the energy $E_{N_\alpha+0.5}=E_{N_\alpha}+14$ MeV. It was suggested that
the number of short range links between $\alpha$-clusters in a nucleus consisting of $ N_{\alpha }$ $\alpha$-clusters equals $3(N_{\alpha }-2)$. Then the nuclear force energy
must be calculated as follows
\begin{equation}
E^{nuc}_{N_\alpha}=N_{\alpha }\varepsilon^{nuc}_{\alpha }+3(N_{\alpha }-2)\varepsilon^{nuc}_{\alpha \alpha },
\label{2}
\end{equation}
where the energy of nuclear force of one
$\alpha $-cluster $ \varepsilon _{\alpha }^{nuc}=\varepsilon _{\alpha }+\varepsilon _{\alpha
}^{C}=29.060$ MeV, $\varepsilon _{\alpha }^{C}$=0.764 MeV. The value $\varepsilon^{nuc}_{\alpha \alpha }=\varepsilon _{\alpha \alpha }+\varepsilon^{C}_{\alpha
\alpha }=4.350$ MeV and $\varepsilon^{C}_{\alpha \alpha }$=1.925 MeV. The parameters were obtained from analysis of the lightest nuclei\cite{3}. To explain (1) one has to suggest that the long range part of the Coulomb interactions between remote $\alpha$-clusters is compensated with the surface tension. Then with applying isospin invariance of nuclear force the empirical values of the Coulomb energy, the energy of surface tension and the empirical values of the distance of the position of the last alpha-cluster $R_{\alpha }$ from the center of the masses of the remote $N_{\alpha }-4$ $\alpha $-clusters for the symmetrical nuclei were found.  The formula for $R_\alpha$
for $Z\geq$ 24  was obtained
\begin{equation}
R_{\alpha }=2.168(N_{\alpha }-4)^{1/3}.
\label{3}
\end{equation}
The following formula for Coulomb radius $R_C$ for the nuclei with $Z, Z_1\geq$ 10 was found
\begin{equation}
R_{C}=1.869N_{\alpha }^{1/3}.
\label{4}
\end{equation}
The charge sphere of the radius $R_{C}$ has the Coulomb
energy 3/5$Z^{2}e^{2}/R_{C}$, which after simplifying
is calculated as the following function\cite{3} of the number $N_\alpha$
\begin{equation}
E^C_{N_\alpha}=1.848(N_{\alpha})^{5/3}.
\label{5}
\end{equation}
The surface tension energy for the nuclei with $Z,Z_1\geq 10$
\begin{equation}
E^{ST}_{N_\alpha}=1+0.471\sum^{N_\alpha}_{6}R_C^2.
\label{6}
\end{equation}
Then the binding energy of a symmetrical nucleus is calculated as follows
\begin{equation}
E_{N_\alpha}=E^{nuc}_{N_\alpha}+E^{ST}_{N_\alpha}-E^{C}_{N_\alpha}. \label{7}
\end{equation}
The Eq. (7) gives rms deviation $\delta$=4 MeV from known today experimental binding energies of symmetrical nuclei with $Z\leq 50$.

The binding energy of excess $nn$-pairs $E_{N_{nn}}=\sum_{1}^{N_{nn}}\epsilon^{sep}_{i_{nn}}$ where $\epsilon^{sep}_{i_{nn}}$ is separation energy of $i^{th}$ $nn$-pair in a stable nucleus, was approximated with
the following function of the number of excess nn-pairs $N_{nn}$\cite{1}
\begin{equation}
E_{N_{nn}}=(21.93-0.762N_{nn}^{2/3})N_{nn}.
\label{8}
\end{equation}
The validity of Eq. (8) is approved by agreement of the values $E_{N_{nn}}$ and the values $E_{exp}-(E^{nuc}_{N_\alpha}+E^{ST}_{N_\alpha}-E^{C}_{N_\alpha})$ where $E_{exp}$ is experimental binding energies of stable and beta -stable isotopes\cite{4}, see Fig.1(a).
\begin{figure}[th1]
\centerline{\psfig{file=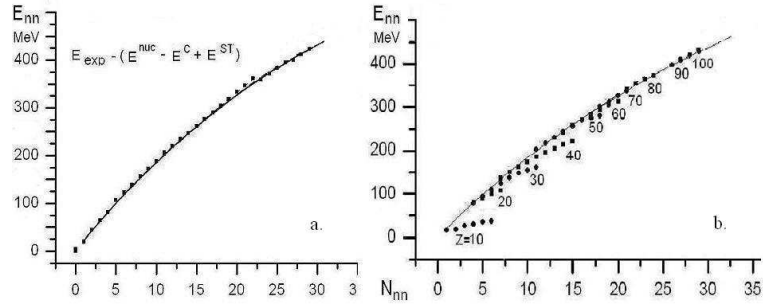,width=10cm}}
\vspace*{8pt}
\caption{Binding energy of excess neutron pairs. a. Square dots indicate the values $E_{exp}-(E^{nuc}+E^{ST}-E^{C})$ for the 41 stable isotopes $A^{st}$ with $Z$=10, 20, 30, 40, 50, 60, 70, 80, 90 è 100. The line is the function $E_{N_{nn}}$ (8). b. The function $E_{N_{nn}}$ (8) is given in comparison with the values $E_{exp}-(E^{nuc}+E^{ST}-E^{C})$ for the isotopes $A \geq A_{max}^{st}$, where $A^{st}_{max}$ is the heaviest stable isotope.}
\end{figure}
The isotopes heavier than the stable ones have the separation energies of the $nn$-pairs smaller than it is implied by (8), see Fig.1(b). A suggestion was made\cite{1} that other $nn$-pairs, which do not find place in the core, come out on the surface of the core. Then the binding energy of $N_{nn}$ $nn$-pairs in an isotope $A(Z,N)$ is calculated as
\begin{equation}
E_{N_{nn}}=E_{N^{stmax}_{nn}}+\sum^{N_{nn}}_{N^{stmax}_{nn}+1}{\epsilon^{sep}_{i_{nn}}},
\label{9}
\end{equation}
where $E_{N^{stmax}_{nn}}$ is the binding energy  of the $N^{stmax}_{nn}$ $nn$-pairs of the heaviest stable isotope and it is calculated by (8). For the other $nn$-pairs the sum of separation energies is used.

Thus, the binding energies of all stable and $\beta $-stable nuclei including the unstable vicinity\cite{3} and the heavier isotopes\cite{1} are calculated with an average deviation from experimental data in a few MeV by the following formula
\begin{equation}
E=E_{N_\alpha}+E_{N_{nn}};E_1=E_{N_\alpha+0.5}+E_{N_{nn}}+\epsilon_n,
\label{10}
\end{equation}
where $\epsilon_n$ is the binding energy of the excess single neutron, which is estimated as the half of the energy of the $N_{nn}+1^{th}$ $nn$-pair.
\subsection{Representation of nucleus as a core (liquid drop) and a molecule on its surface}
The representation of nucleus allows one to widen the number of isotopes to be described to those with $N\geq Z$. The binding energies $E$ and $E_{1}$ of the nuclei $A(Z,N)$ and $A_{1}(Z_{1},N+2)$ are calculated as follows\cite{1}
\begin{equation}
E=E_{N_{\alpha }^{ml}}+E_{core}-E^C_{N^{ml}_{\alpha }N^{core}_\alpha
};E_{1}=E_{N^{ml}_\alpha +0.5}+E_{core}-E^{C}_{N^{ml}_\alpha+0.5N^{core}_\alpha },  \label{11}
\end{equation}
where $E_{N_{\alpha }^{ml}}$ and $E_{N^{ml}_\alpha +0.5}$ are the
binding energies of the nuclei with the number $N_{\alpha }^{ml}$ and $N^{ml}_\alpha +0.5$
$\alpha$-clusters. For example $E_{3}=E_{^{12}\mathrm{C}}$ and $E_{3.5}=E_{^{15}\mathrm{N}}$, $ E_{N_{\alpha}^{ml}N_{\alpha }^{core}}^{C}$ is the energy of the Coulomb interaction between the peripheral molecule and the core
\begin{equation}
E^{C}_{N_{\alpha}^{ml}N_\alpha^{core}}=2N_{\alpha }^{ml}2N_{\alpha
}^{core}e^{2}/R_{\alpha }.
\label{12}
\end{equation}
The core binding energy is calculated in accordance with the representation as
it follows
\begin{equation}
E_{core}=E_{N_{\alpha }^{core}}+6\epsilon _{\alpha \alpha }+E_{N_{nn}},
\label{13}
\end{equation}
where $E_{N_{\alpha }^{core}}$ is calculated by (7) on the number $N_{\alpha }^{core}$. Six short range links come to keep the total number of short range links in the whole nucleus equal to $3(N_\alpha -2)$, because the total number of links in the two separate objects is on six links less: $3(N_\alpha^{core}-2)+ 3(N_\alpha^{ml}-2)=3N_\alpha-12$.

In the calculation of surface tension energy for the core $\alpha$-clusters $E^{ST}_{N^{core}_\alpha}$ a simple approximation of (6) is used \cite{1,3}
\begin{equation}
E^{ST}_{N^{core}_\alpha}=(N_{\alpha }+1.7)(N_{\alpha }-N^{ml}_{\alpha })^{2/3}  \label{14}
\end{equation}
at $N_{\alpha}^{ml}$=5 for the nuclei with $Z \leq $36 and  $N_{\alpha}^{ml}=4$ for the nuclei with $Z \geq$ 36 with the biggest deviation between the functions (14) and (6) in 2 MeV at $Z$=36.  This allows one to put the surface tension energy in dependence on the number of core $\alpha$-clusters.

From fitting experimental nuclear binding energies with the formula (11)
one can find the number of alpha-clusters in the molecule $N^{ml}_\alpha $.
The analysis shows that in isotopes of one $Z, (Z_1)$ with growing $A, (A_1)$ the number $N^{ml}_\alpha$ decreases in most of the cases to $N^{ml}_\alpha$=3 and the specific density of core binding energy $\rho $ slowly increases to reach its saturation value in stable and beta stable isotopes. It was shown on examples given in Table\cite{1}, that separation energy of an excess neutron pair is not always equal to its binding energy. If at separation of one $nn$-pair the number of $N^{ml}_\alpha$ decreases ($N^{core}_\alpha$ correspondingly increases), $\epsilon^{sep}_{i_{nn}}$ is bigger than $\epsilon_{i_{nn}}=E_{i_{nn}}-E_{i_{nn}-1}$. In the isotopes heavier than stable ones the core stays unchanged and other neutron pairs gather  on the surface of the core. In that case the binding energies of the $nn$-pairs are equal to their separation energies. In this representation binding energies for some heavy isotopes $A$ have been predicted, for the cases when the experimental  binding energy of $A_1$ are known, and in reverse\cite{1}.

The obtained value $N^{ml}_\alpha$ allows one to calculate the radii. There are four simple formulas to calculate radii\cite{1}.  It was shown that charge radius is defined by the number of alpha-clusters in the nucleus rather than by the mass number $A$. Charge radius slowly decreases with growing $A(A_1)$ due to decreasing the number $N^{ml}_\alpha$ and after reaching stability stays unchanged. The matter radius grows with the number of excess neutrons. The matter radius of a stable nucleus is expected to be equal to its charge radius. From fitting the experimental radii the charge radius of an alpha-cluster of core is found to be $r_\alpha$=1.595 fm and the radius of the volume occupied by one proton/neutron in the alpha-cluster is defined by the radius $r_{n/p}$=0.954 fm and radius of one excess neutron in a $nn$-pair is $r_{n}$=0.796 fm. The calculated values are in a good agreement with the experimental data\cite{5} (known mostly for stable nuclei) with the average deviation in 0.028 fm\cite{1}.

\section{New proofs of validity of the model }

In the present work some new proofs of the model representation have been found. It is well known that unlike even $Z$ nuclei there is only one rare two stable isotopes with odd $Z_1=Z+1$ and the number of excess neutrons is always odd. The finding is that the binding energies of the chain of stable nuclei  $A_1(Z_1,N+2)$ and the corresponding $A(Z,N)$ nuclei are roughly described by the following formula
\begin{equation}
E=(N_\alpha-3)v_{\alpha }\rho+E_{ml}-Z_{ml}2(N_\alpha-3)e^{2}/R_{\alpha },
\label{15}
\end{equation}
where in the sum the first summand is the core binding energy, the volume of one core $\alpha$-cluster $v_\alpha=4/3\pi r_\alpha^3$ fm$^3$ is defined by its charge radius $r_\alpha=1.595$ fm, which is obtained from independent analysis of experimental charge radii; the specific density of the core binding energy $\rho=2.55$MeV/fm$^3$  is the only fitting parameter here; the second summand  is the binding energy of $^{12}$C $E_{ml}$=92 MeV (for odd $Z_1$ $^{15}$N $E_{ml}$= 115 MeV) and the third summand is the Coulomb energy of interaction between the outside molecule of $Z_{ml}$=6 ($Z_{ml}$ =7) and the core, where the distance between the objects $R_\alpha$ is taken as: $i$ the empirical values of $R_\alpha$ in the case of $Z<24$, see \cite{3}, and the calculated values (3) for the other $Z\geq24$; $ii$ $R_\alpha$ is equal to the sum of the radii of these two objects where $R_{core}=r_\alpha (N_\alpha-3)^{1/3}$ and $R_{ml}=r_{^4\rm He}3^{1/3}$ ($R_{ml}=r_{^4\rm He}3.5^{1/3}$) with $r_{^4\rm He}$=1.71 fm (an experimental radius of the nucleus $^4$He).
For $i$ the value $\rho$=2.56 MeV/fm$^3$ with rms deviation from stable nuclei with $Z,Z_1\leq83$ $\delta$=17 MeV and for $ii$ $\rho$=2.54 MeV/fm$^3$ with $\delta$=15 MeV. In Fig. 2 the binding energy (15) of stable and beta-stable nuclei for the isotopes $A_1(Z_1,N+2)$ and $A(Z,N)$ are shown in comparison with the experimental data. At the graph scale the corresponding to $i$ and $ii$ lines are not separated. Eq. (15) means that stability is provided by the portion of the binding energy in $v_{\alpha }\rho$=43.34 MeV per one core $\alpha$-cluster.

The binding energies of the chain $A_1(Z_1,N+2)$, $A(Z,N)$ of the beta-stable isotopes with $Z\geq$84 are well described with two outside molecules, these are $^4$He and $^{12}$C ($^{15}$N) and bigger ones. The formula to calculate binding energy in the case of two molecules outside the core is (8)\cite{1}.

\begin{figure}[th2]
\centerline{\psfig{file=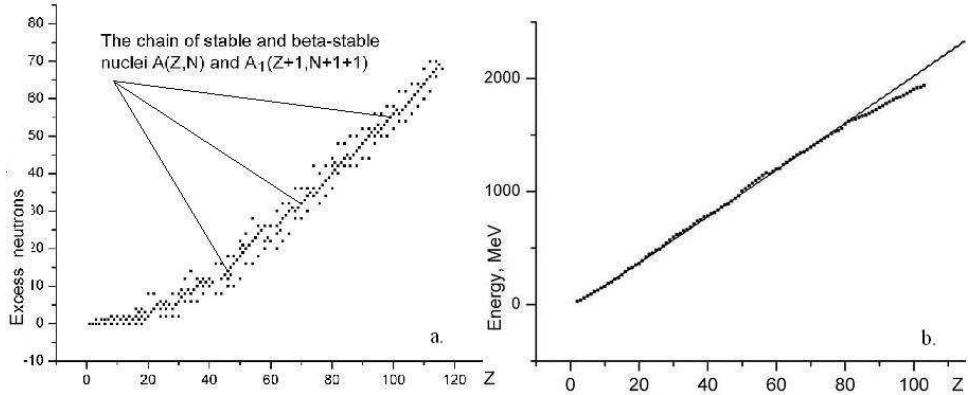,width=13cm}}
\vspace*{8pt}
\caption{a. Numbers of excess neutrons $\Delta N$ for the chain of stable and beta-stable nuclei $A(Z,N)$ and $A_1(Z_1,N+2)$. Also the values $\Delta N$ for the lightest and the heaviest even $Z$ stable and beta-stable isotopes are given. b. Binding energies of stable and beta-stable nuclei $A(Z,N)$ and $A_1(Z_1,N+2)$. The line indicates calculated binding energies  (15) in comparison with experimental data.}
\end{figure}

 The number of excess neutrons $\Delta N=2N_{nn}$ in case $A(Z,N)$ and $\Delta N=2N_{nn}+1$ in case of $A_1(Z_1,N+2)$. The binding energy of core  is calculated as $E_{core}=N^{core}_\alpha v_{\alpha}\rho=43.34 N^{core}_\alpha$. Then Eq. (13) is rewritten as follows
\begin{equation}
E_{N_{nn}}=43.34N^{core}_\alpha-(E_{N^{core}_\alpha}+6\epsilon_{\alpha \alpha}),
\label{16}
\end{equation}
where $N^{core}_\alpha=N_\alpha$-3. The value $\Delta N$ is obtained as solution of two equations (8) and (16) at the condition that the number $N_{nn}$ is integer. The numbers of excess neutrons calculated in the representation that they fill out the core are in an agreement with the actual ones, see Fig.3.

\begin{figure}[th3]
\centerline{\psfig{file=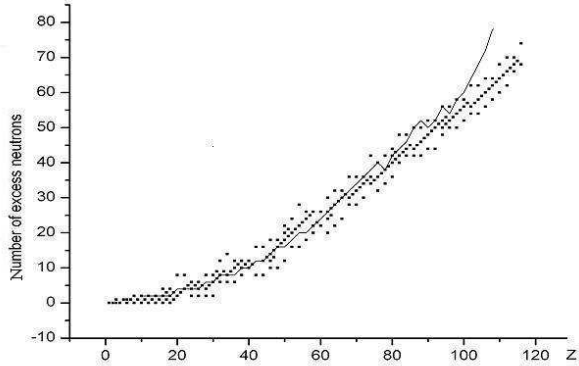,width=8cm}}
\vspace*{8pt}
\caption{The number of excess neutrons in stable and beta-stable nuclei $À(Z,N)$ and
$A_1(Z_1,N+2)$. The line is $\Delta N=2N_{nn}$ for $A$ and $\Delta N=2N_{nn}+1$ for $A_1$
where $N_{nn}$ is the solution of two
equations (8) and (16) at $N^{core}_\alpha=N\alpha-3$ for $Z, Z_1\leq 78$,  $N^{core}_\alpha=N_\alpha-4$ for
79$\leq Z, Z1 \leq$89, $N^{core}_\alpha=N\alpha-5$ for
90$\leq Z, Z1 \leq$96, $N^{core}_\alpha=N\alpha-6$ for
$Z, Z1 \geq$96.}
\end{figure}

From the figure it is seen that for stable nuclei with $Z \leq 78$ the outside molecule in three alpha-clusters (three and half in case odd $Z_1$)  provides a good describing the actual numbers of the excess neutrons.

\section{Conclusions}

The binding energies and radii of the nuclei with $N\geq Z$ are described in the framework of the alpha-cluster model. The number of excess neutron pairs to fill out the core in stable nuclei is determined by the size of the core defined by the number of core $\alpha$-clusters $N^{core}_\alpha=N_\alpha-3$ and by the saturation value of the specific density of the core binding energy $\rho$ = 2.55 MeV/fm$^3$.

\end{document}